\documentclass{aastex63}
\usepackage{epstopdf}

\submitjournal{ApJ}

\begin{document}

\title{Small-scale \textbf{Bright Blobs}  Ejected from a Sunspot Light Bridge}
\correspondingauthor{Yajie Chen}
\email{chenyajie@pku.edu.cn}

\correspondingauthor{Hui Tian}
\email{huitian@pku.edu.cn}

\author{Fuyu Li}
\affiliation{School of Earth and Space Sciences, Peking University, 100871 Beijing, China}

\author{Yajie Chen}
\affiliation{School of Earth and Space Sciences, Peking University, 100871 Beijing, China}
\affiliation{Max-Planck Institute for Solar System Research, Justus-von-Liebig-Weg 3, D-37077 G\"{o}tingen, Germany}

\author{Yijun Hou}
\affiliation{Key Laboratory of Solar Activity, National Astronomical Observatories, Chinese Academy of Sciences, Beijing 100101, China}

\author{Hui Tian}
\affiliation{School of Earth and Space Sciences, Peking University, 100871 Beijing, China}
\affiliation{Key Laboratory of Solar Activity, National Astronomical Observatories, Chinese Academy of Sciences, Beijing 100101, China}

\author{Xianyong Bai}
\affiliation{Key Laboratory of Solar Activity, National Astronomical Observatories, Chinese Academy of Sciences, Beijing 100101, China}

\author{Yongliang Song}
\affiliation{Key Laboratory of Solar Activity, National Astronomical Observatories, Chinese Academy of Sciences, Beijing 100101, China}

\begin{abstract}
Light bridges (LBs) are bright lanes that divide an umbra into multiple parts in some sunspots. Persistent oscillatory bright fronts at a temperature of $\sim$$10^5$ K are commonly observed above LBs in the 1400/1330 \AA~passbands of the Interface Region Imaging Spectrograph (IRIS). Based on IRIS observations, we report small-scale bright blobs from the oscillating bright front above a light bridge.  Some of these blobs reveal a clear acceleration, whereas the others do not. The average speed of these blobs projected onto the plane of sky is $71.7\pm14.7$ km s$^{-1}$, with an initial acceleration of $1.9\pm1.3$ km s$^{-2}$. These blobs normally reach a projected distance of 3--7 Mm from their origin sites. From the transition region images we find an average projected area of $0.57\pm0.37$ Mm$^{2}$ for the blobs. The blobs were also detected in multi-passbands of the Solar Dynamics Observatory, but not in the H$\alpha$ images. These blobs are likely to be plasma ejections, and we investigate their kinematics and energetics. Through emission measure analyses, the typical temperature and electron density of these blobs are found to be around $10^{5.47}$ K and $10^{9.7}$ cm$^{-3}$, respectively. The estimated kinetic and thermal energies are on the order of $10^{22.8}$ erg and $10^{23.3}$ erg, respectively. These small-scale blobs appear to show three different types of formation process. They are possibly triggered by induced reconnection or release of enhanced magnetic tension due to interaction of adjacent shocks, local magnetic reconnection between emerging magnetic bipoles on the light bridge and surrounding unipolar umbral fields, and plasma acceleration or instability caused by upward shocks, respectively.
\end{abstract}


\section{Introduction} \label{sec:intr}

Bright lanes that divide an umbra into multiple parts, called light bridges (LBs), are among the most prominent bright structures in many sunspots. The magnetic field at LBs is often much weaker and more inclined compared to that of the umbra \citep{Leka1997, Shimizu2009, Toriumi2015b,Yuan2016, Feng2020}. High-resolution observations have shown dark substructures within light bridges, such as central dark lanes with upflows \citep{Berger2003, Rouppe2010} and dark knots with downflows \citep{Zhang2018}. These small-scale structures are likely formed as a result of the magnetoconvection in sunspots \citep{Sch2006}. Recent observations and simulations have revealed possible signatures of weakly twisted magnetic field and flux emergence in LBs \citep{Louis2015, Tian2018a, Toriumi2015b, Toriumi2015a, Song2017,YangHeesu2019, Yuan2016, Bai2019, Lim2011}.

The most prominent phenomenon above LBs revealed from chromospheric observations is the recurrent surge-like activity, which was reported as H$\alpha$ surges, light wall oscillations, plasma ejections, or chromospheric jets \citep{Roy1973, Asai2001, Bharti2007, Bharti2015, Shimizu2009, Hou2016a, Pontieu2017, Yuan2016, Song2017}. External disturbances such as flares and falling materials could change the amplitudes of the light wall oscillations \citep{Hou2016b, Yang2016}. Some of these jets have very high speeds ($\sim$100 km s$^{-1}$) and were found at locations of strong electric current \citep{Louis2014, Robustini2016, YangXu2019, Lim2020}, suggesting that they are likely driven by magnetic reconnection between small-scale magnetic bipoles at LBs and the surrounding unipolar umbral field. Meanwhile, low-speed surges ($\sim$15 km s$^{-1}$) with a stable oscillating period of a few minutes have also been reported, and suggested to be driven by p-mode or magnetoacoustic waves leaking from the underlying photosphere \citep{Yang2015, Zhang2017}. Based on observations of the Goode Solar Telescope \citep[GST,][]{Cao2010} and the Interface Region Imaging Spectrograph \citep[IRIS,][]{De2014}, \citet{Tian2018} found strong evidence of two types of surge-like activity above LBs: persistent up-and-down motions associated with the upward leakage of magnetoacoustic waves from the photosphere, and sporadic high-speed jets triggered by intermittent magnetic reconnection. Based on imaging observations of IRIS, \cite{Hou2017} also reached a similar conclusion.

Bright fronts ahead of LB surges are commonly observed in transition region (TR) images \citep{Bharti2015,Yang2015}. It is likely that the bright fronts ahead of the LB surges is heated to TR temperatures either by shocks or through compression \citep{Morton2012,Bharti2015,Zhang2017}. In this paper, we report small-scale bright blobs at typical TR temperatures from the bright front of surges above a light bridge, which is rarely observed in the past. The existence of these blobs indicates more complex physical processes above the LB. We perform a detailed investigation of their kinematics and energetics, and discuss their possible formation mechanisms.

\section{Observations} \label{sec:data}
The observations were performed from 05:17:13 UT to 06:30:18 UT on 2014 Oct 25. The IRIS pointed to a coordinate of (257$''$, -318$''$), targeting at the largest sunspot group of solar cycle 24 located in NOAA Active Region (AR) 12192. We used the level 2 IRIS data, which has been corrected through dark current subtraction, flat field, geometrical, and orbital variation corrections. Slit-jaw images (SJIs) of IRIS in the filter of 1400 \AA~were used, and the cadence was $7.3$ s. The field of view (FOV) of SJIs was 119$''$ $\times$ 119$''$, with a spatial pixel size of 0.$^{\prime\prime}$166. The roll angle of the spacecraft was $45\arcdeg$. Since the slit failed to catch any blobs investigated in this paper, we did not use the spectral data.

We also analyzed the data taken simultaneously by the Atmospheric Imaging Assembly \citep[AIA,][]{Lemen2012} on board the Solar Dynamics Observatory (SDO) to study the temperature of the blobs. The cadence of AIA observations was 12~s in the 94, 131, 171, 193, 211, 304, 335 \AA~passbands, and 24~s in the 1600 \AA~passband. The pixel size of the AIA images was $\sim$0.$^{\prime\prime}$613. The AIA images in different passbands were coaligned using the IDL routine aia$\_$prep.pro available in Solar Software. The AIA and IRIS images were then coaligned by comparing some commonly observed bright features in the AIA 1600 \AA~images and the IRIS/SJI 1400 \AA~images.

Moreover, the 1-m New Vacuum Solar Telescope \citep[NVST,][] {Liu2014,Yan2020} also observed this sunspot from 04:59:20 UT to 05:59:15 UT. The H$\alpha$ core images of NVST were used to examine the chromospheric response of the blobs. The bandwidth of the H$\alpha$ filter was 0.25 \AA. These H$\alpha$ core images have an image scale of 0.$^{\prime\prime}$164 pixel$^{-1}$, a cadence of $12~s$, and a FOV of $152'' \times 152''$. We coaligned the H$\alpha$ images and SJI 1400 \AA~images by matching the commonly observed LB and some localized dynamic features.

\section{Data Analysis and Results\label{subsec:analysis and results}}

\subsection{Kinematics of the ejections\label{subsec:ejections}}
\begin{figure}[ht!]
\plotone{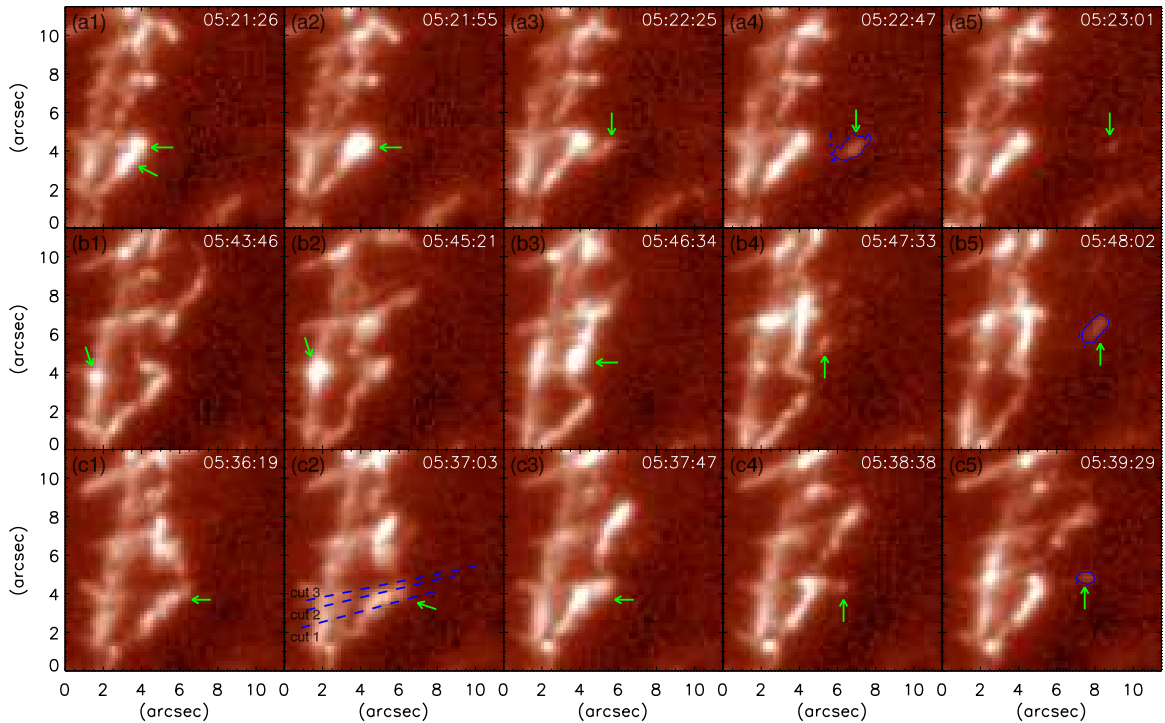}
\caption{Temporal evolution of three bright blobs in SJI 1400 \AA~images. Panels (a1)-(a5), (b1)-(b5), and (c1)-(c5) are for the blobs 1, 6, and 4 listed in Table 1, respectively. The blue contours in panels (a4), (b5), and (c5) outline the locations of the blobs. The blue dashed lines in panel (c2) mark three cuts that we used to produce space-time diagrams in Figure~\ref{fig:stplotfronts}. The green arrows highlight the blobs and some brightenings related to the origin of these blobs. The images are shown in logarithmic scale. (An animation of this figure is available.) \label{fig:merge}}
\end{figure}

\cite{Bharti2015} and \cite{Yang2015} analyzed this dataset and identified an oscillating bright front above the LB from the TR images. The bright front is visible from the 1400 \AA~images shown in Figure \ref{fig:merge}. It exhibits up-and-down oscillatory motions during the entire observation period. Besides this oscillating bright front, we also identified more than ten small-scale intermittent bright blobs away from the leading edge. \textbf{The blobs are better seen in the movies, and we pointed out each blob by a green arrow in Movie 1. These bright blobs may correspond to plasma ejections, and we cannot rule out the the possibility of heating fronts due to the absence of spectral data for the bright blobs. Considering both network jets and penumbral microjets, which could be caused by heating fronts \citep{Pontieu2017, Chen2019, Rouppe2019, Esteban2019}, usually perform elongated structures, and the bright blobs we report in this study exhibit circular shape, so they are more likely mass ejections.}
 
 The lifetimes of these blobs are mostly 30-100 s from the 1400 \AA~image sequence. Most of these blobs appear to be initiated from the bright front, and the rest might be ejected from the LB base. We identified 13 blobs that could be unambiguously identified and tracked from the 1400 \AA~images. All of these blobs show only upward propagation and then disappear. No obvious signs of downward motions were found. These bright features were noticed by \cite{Bharti2015}, but no detailed analysis was performed. Observational details of these bright blobs are listed in Table 1.

\begin{deluxetable*}{cccccccccccc}
\tablenum{1}
\tablecaption{Observational details and measured physical parameters of the bright blobs. \label{tab:1}}
\tablewidth{0pt}
\tablehead{
\colhead{No.} & \colhead{Time} & \colhead{$D$} & \colhead{$v$} & \colhead{$a$} & \colhead{$S$} &\colhead{log$_{10}$~EM}&\colhead{log$_{10}$~$T$}& \colhead{log$_{10}$~$n_e$} &\colhead{log$_{10}$~$E_k$}&\colhead{log$_{10}$~$E_t$}&\colhead{Type}\\
\colhead{} & \colhead{(UT)} & \colhead{$km$} & \colhead{km s$^{-1}$} & \colhead{km s$^{-2}$} & \colhead{Mm$^2$}&\colhead{cm$^{-5}$} &\colhead{K}& \colhead{cm$^{-3}$} &\colhead{erg}&\colhead{erg}
}
\startdata
1 & 05:22:10 - 05:23:09 & 3980 & 68.0  & 0.0 & 0.85 &27.35&5.45&9.69&23.25&23.66&\uppercase\expandafter{\romannumeral1}\\
2 & 05:33:31 - 05:34:00 & 2291 &  78.2  &  3.4 & 0.06 &27.15&5.49&9.88&21.84&22.16&\uppercase\expandafter{\romannumeral1}\\
3 & 05:37:55 - 05:39:15 & 6874 &85.3   & 2.2 & 0.45 &27.17&5.45&9.67&23.02&23.22&\uppercase\expandafter{\romannumeral1}\\
4 & 05:39:08 - 05:39:45 & 2533& 69.2  & 3.4 & 0.24 &27.20&5.50&9.75&21.52&22.96&\uppercase\expandafter{\romannumeral3}\\
5 & 05:41:20 - 05:42:11 & 3136 & 61.2  &  1.1 & 0.31 &-&-&-&-&-&\uppercase\expandafter{\romannumeral1}\\
6 & 05:47:11 - 05:48:10 & 4101 & 70.0 & 0.0 &  0.68 &27.15&5.43&9.62&23.06&23.41&\uppercase\expandafter{\romannumeral2}\\
7 & 05:49:30 - 05:50:07 & 3136 &85.6  & 2.3 & 0.36 &27.18&5.48&9.70&22.90&23.13&\uppercase\expandafter{\romannumeral1}\\
8 & 05:51:42 - 05:52:48 & 6151 & 93.4 & 3.4 & 1.27 &27.21&5.47&9.58&23.68&23.82&\uppercase\expandafter{\romannumeral1}\\
9 & 05:59:45 - 06:00:44& 5186 & 88.6  & 1.1 &1.13 &27.50&5.50&9.74&23.71&23.93&\uppercase\expandafter{\romannumeral1}\\
10 & 06:07:56 - 06:09:01 & 4704&71.4 & 3.4 & 0.88 &27.30&5.49&9.66&23.29&23.69 &\uppercase\expandafter{\romannumeral3}\\
11 & 06:12:19 - 06:13:40 & 3256& 40.4& 2.3 & 0.32 &27.12&5.46&9.68&22.15&23.01 &\uppercase\expandafter{\romannumeral1}\\
12 & 06:23:11 - 06:23:47 & 2050&56.0 & 1.1 &  0.25 &27.13&5.45&9.72&22.30&22.87&\uppercase\expandafter{\romannumeral3} \\
13 & 06:27:12 - 06:28:55 & 6633& 64.7& 1.1 & 0.59 &-&-&-&-&- &\uppercase\expandafter{\romannumeral3}\\
\enddata
\tablecomments{The second column gives the time ranges during which the blobs appear in SJI 1400 \AA ~images. Other columns list the propagation distance ($D$), velocity ($v$), initial acceleration ($a$), area ($S$), emission measure (log$_{10}$~EM), temperature (log$_{10}$~$T$), electron density (log$_{10}$~$n_e$), kinetic energy (log$_{10}$~$E_k$), and thermal energy (log$_{10}$~$E_t$) for the 13 bright blobs. Since blobs 5 and 13 cannot be unambiguously identified from the AIA channels, their energies and emission measures are not calculated. The last column shows the types of their possible formation processes.}
\end{deluxetable*}

There appear to be three types of bright blobs that are possibly associated with different formation processes. Figure~\ref{fig:merge} shows the IRIS image sequences for  three blobs, each falling into one of the three types. In the first example (Figure~\ref{fig:merge}(a1)-(a5), blob 1 in Table~\ref{tab:1}, type \uppercase\expandafter{\romannumeral1}), we see two small-scale brightenings at the bright front before the bright blobs. These two brightenings subsequently merge into a bigger one (Figure~\ref{fig:merge}(a2)), from which an blob is then clearly separated. For most of the eight type I cases, the newly formed features are also brighter. 

The second example (Figure~\ref{fig:merge}(b1)-(b5), blob 6, type \uppercase\expandafter{\romannumeral2}) appears to have a different formation process. Before the formation of the bright blob, there is an evident brightening (Figure~\ref{fig:merge}(b2)) at the LB \citep[similar to the first case studied in][]{Hou2017}. Then the emission of the bright front is enhanced at the location right above this brightening, where a faint jet-like structure appeares to connect the LB and the bright front. Later on, a bright blobs is ejected away from the bright front. We only found one such case (type II). 

\begin{figure}
   \centerline{\includegraphics[width=0.6\textwidth]{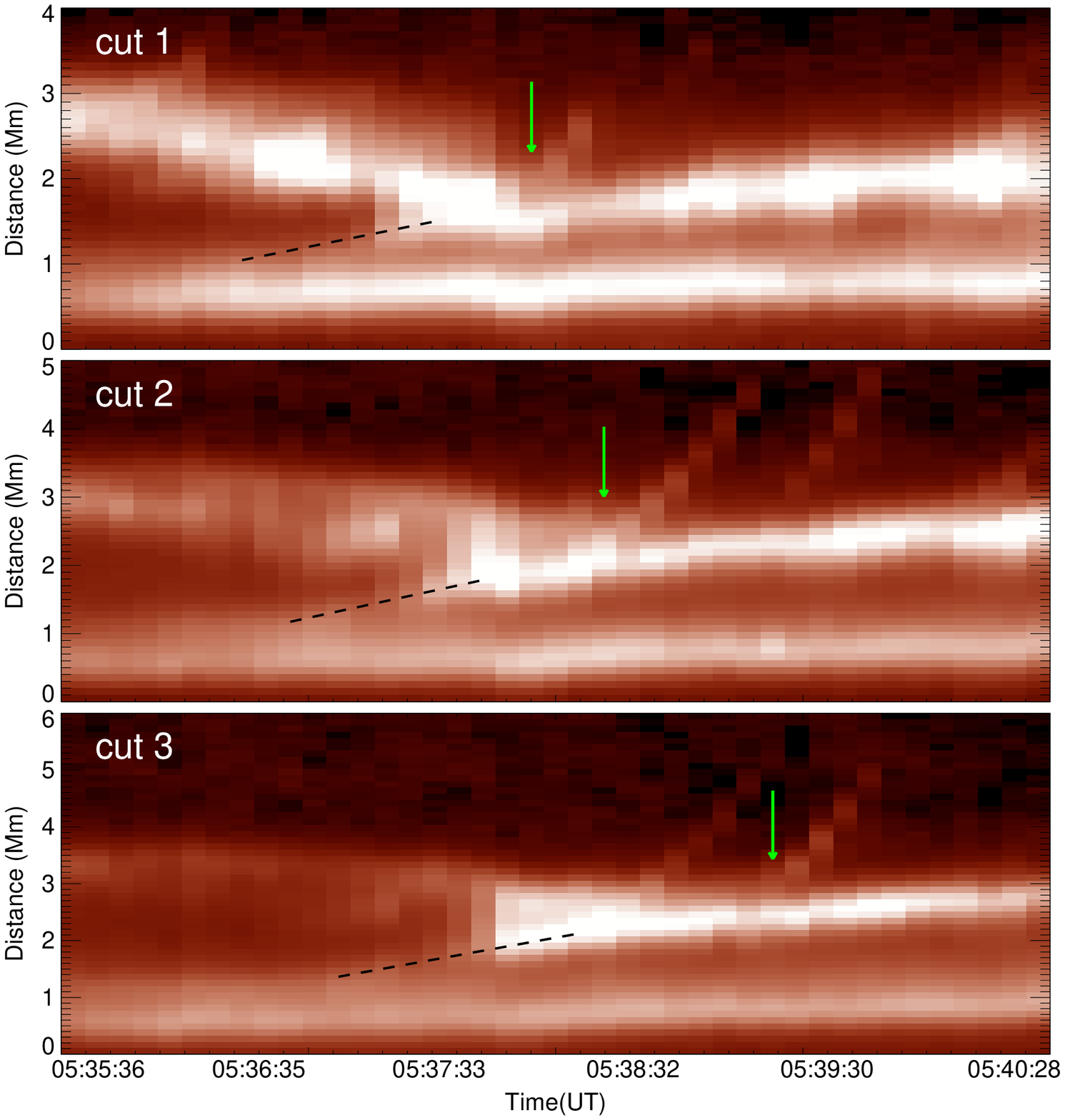}}
\caption{Space-time diagrams along the cuts shown in Figure \ref{fig:merge} (c2). The black dashed lines indicate the rising motion of some materials from the LB. The green arrows indicate the apparent origin sites of bright blobs. \textbf{The dashed black lines indicate the speeds of the bright edges, that is around $1 Mm~s^{-1}$.}
 \label{fig:stplotfronts}}
\end{figure}

The third example (Figure~\ref{fig:merge}(c1)-(c5), blob 4, type \uppercase\expandafter{\romannumeral3}) is different from the two examples described above. Before the bright blobs forms, neither significant intensity enhancement at the LB nor signature of merging bright points at the bright front is observed. However, we noticed that materials arise from a wide range of the LB, and they merge with the bright front. Then three successive bright blobs appear after 05:37 UT. Figure \ref{fig:stplotfronts} shows the space-time diagrams along different paths (Figure~\ref{fig:merge}(c2)) of the three successive blobs. It is evident that when the rising materials from the LB merge with the bright front from the south to the north, bright blobs successively occur at the different merging locations. There are four such blobs (type III).

\begin{figure}[ht!]
   \centerline{\includegraphics[width=0.6\textwidth]{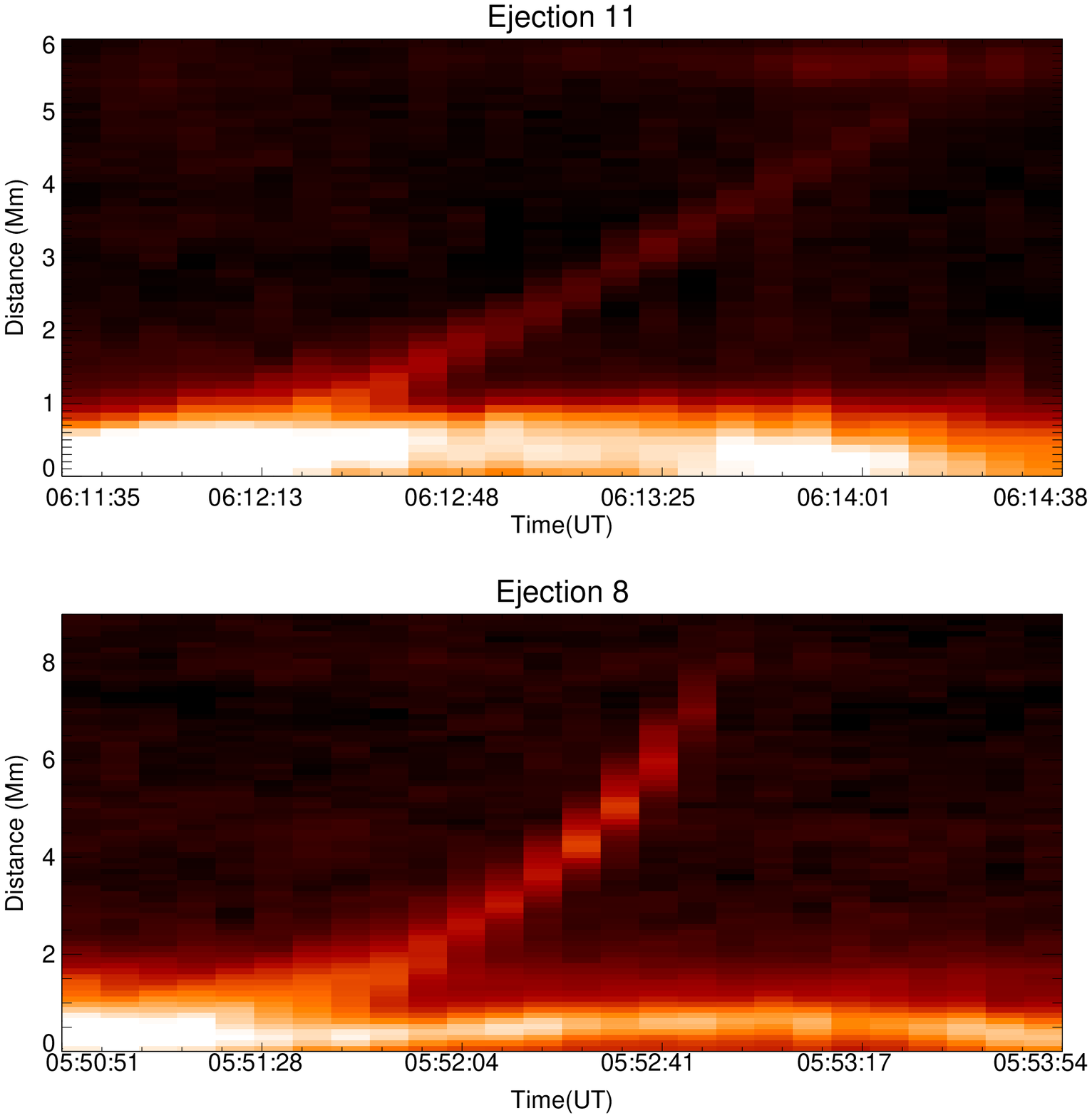}}
\caption{Space-time diagrams along the paths of \textbf{bright blobs} 11 and 8. \label{fig:diagram}}
\end{figure}

To measure the velocities ($v$) and initial accelerations ($a$) of these blobs, we produced space-time diagrams along the paths of each blob. Figure~\ref{fig:diagram} shows two examples. Blob 11 reveals a constant speed, whereas blob 8 has an obvious acceleration during the initial propagation. We calculated the average speed and acceleration for each case. These blobs have speeds of $71.7\pm14.7$ km s$^{-1}$, with accelerations of $1.9\pm1.3$ km s$^{-2}$. For the only one type \uppercase\expandafter{\romannumeral2} blob, its acceleration is 0 km s$^{-2}$. While for most of the other blobs, we see a clear acceleration. We also calculated the propagation distances ($D$) and projected areas ($S$) of the blobs. These blobs normally propagate to a projected distance of 3--7 Mm from the bright front in the 1400 \AA~passband. The projected areas, with an average of $0.57\pm0.37$ Mm$^{2}$, do not reveal a significant change during the propagation. These kinematic parameters of all blobs are listed in Table~\ref{tab:1}.

\subsection{Response in AIA passbands\label{subsec:aia}}

\begin{figure}[ht!]
\plotone{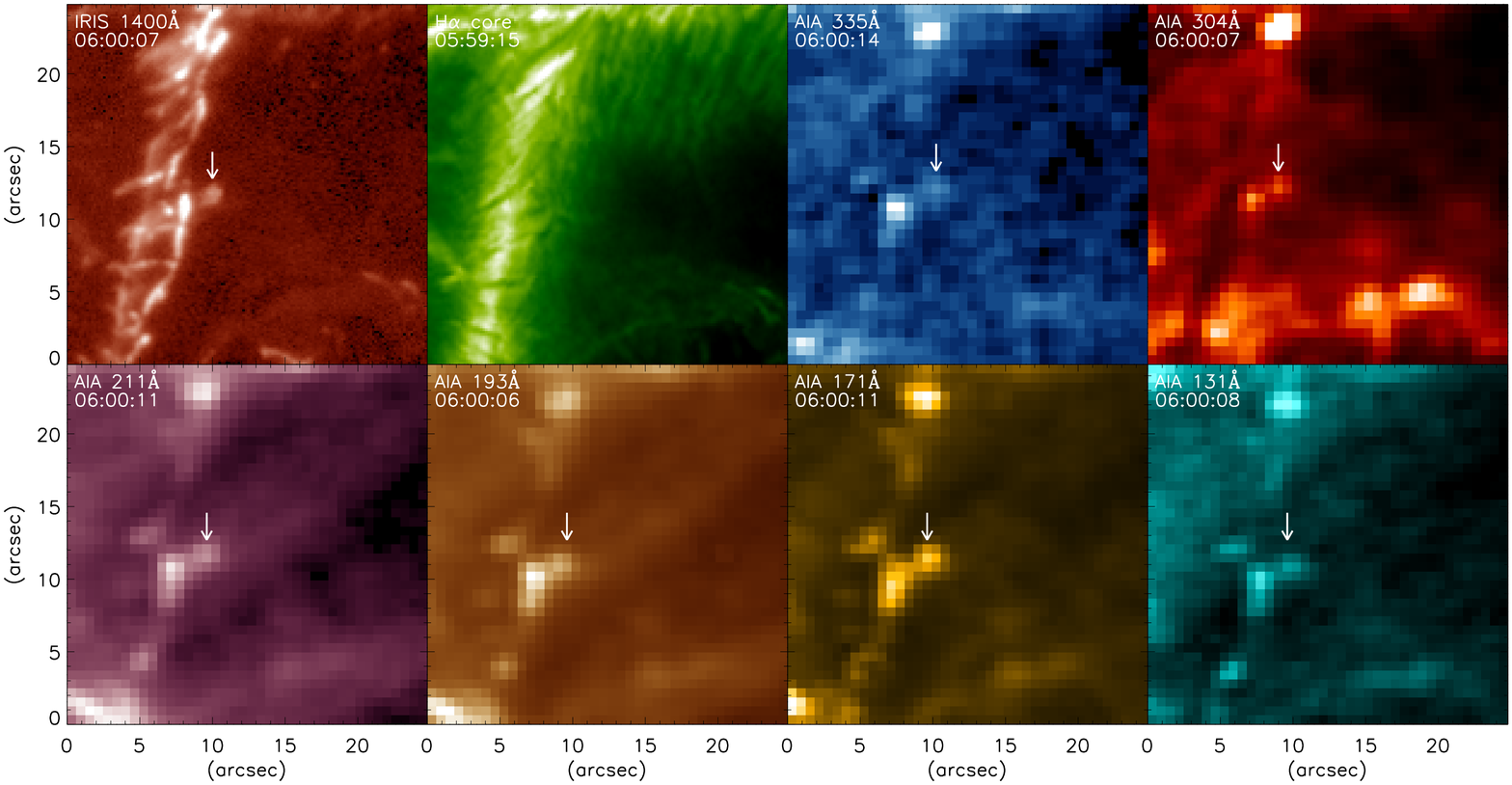}
\caption{A blob (indicated by the white arrows) in the SJI 1400 \AA, NVST H$\alpha$ core, and AIA 335, 304, 211, 193, 171 and 131 \AA~images. An animation of this figure is available online. (An animation of this figure is available.)\label{fig:AIA}}
\end{figure}

Most blobs also show clear signatures in images of all or most of the AIA 131, 171, 193, 211, 304, and 335 \AA~filters, and an example is presented in Figure~\ref{fig:AIA}. Meanwhile, these blobs do not show any obvious signatures in H$\alpha$ core images. The visibility of these blobs in different AIA passbands and SJI 1400 \AA~images suggests either a multi-thermal structure or a TR temperature. It is difficult to determine the temperature structure through a differential emission measure (DEM) analysis since the low-temperature part of the DEM can not be well constrained by observations in the AIA passbands \citep{Del2011,Testa2012}.

\begin{figure}[ht!]
\plotone{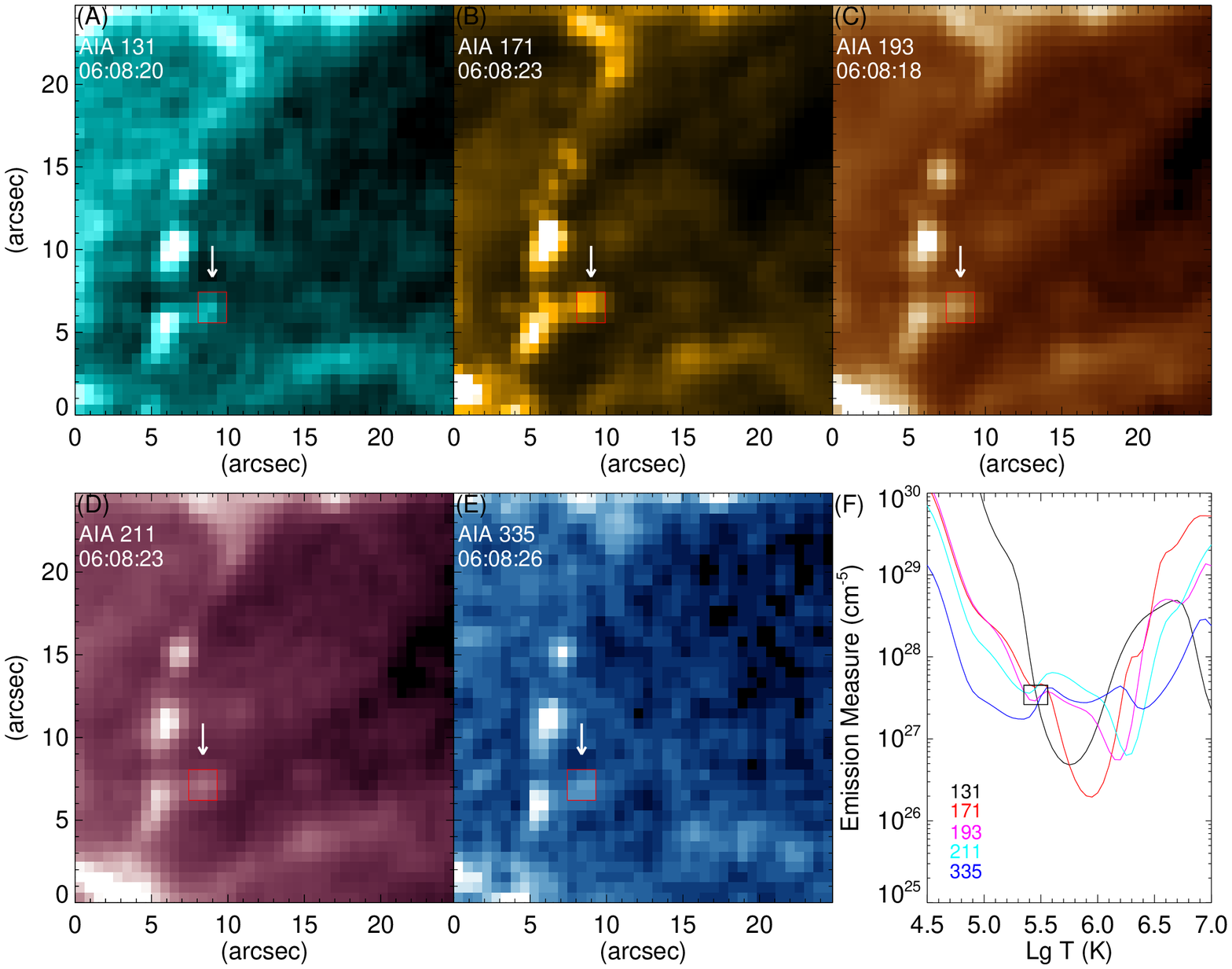}
\caption{(A)-(E) AIA 131, 171, 193, 211 and 335 \AA~images taken around 06:08:24 UT. The white arrow and red square in each of panels (A)-(E) indicate a plasma blob and the region for an EM-loci analysis, respectively. (F) The EM-loci curves for the blob. The small black box shows a region with many crossings of the EM-loci curves.  \label{fig:EMloci}}
\end{figure}

Instead, we calculated the EM-loci (emission measure loci) curves for each plasma blob to determine the possible temperature \citep[e.g.,][]{Del2002, Amy2013, Tian2014}. Figure~\ref{fig:EMloci} shows the EM-loci curves for blob 10. The EM-loci curves of different AIA passbands cross a relatively narrow box around the temperature of $10^{5.5}$ K and emission measure (EM) of $10^{27.25}$ cm$^{-5}$. This temperature and EM are regarded as rough estimates of the plasma temperature and EM of the blob. Two of the plasma blobs (blobs 5 and 13 in Table 1) do not show obvious signatures in most of the AIA EUV passbands, and we could not perform an EM-loci analysis. The possible temperature and EM values for the rest of the identified plasma blobs are shown in Table~\ref{tab:1}. We found that the EM and temperature of the ejected plasma are on average $10^{27.22}$ cm$^{-5}$ and $10^{5.47}$ K, respectively. The EM-loci curves should just be considered as an upper limit of the true emission measure distribution \citep{Tian2014}.

\subsection{Energy estimation\label{subsec:energy}}

Using the EM and temperature estimated from the EM-loci analysis, we can calculate the kinetic and thermal energy for each blob. Under the assumption that the integration length along the line of sight is similar to the projected size, we estimated the electron density ($n_e$) of a blob using Equation~\ref{e1},
\begin{equation}
EM\approx n_e^2L
\label{e1}
\end{equation}
where $L$ is the square root of the area of the blob. The electron densities of the blobs were found to be $10^{9.70\pm0.08}$~cm$^{-3}$.

Assuming a fully-ionized plasma and a typical coronal composition \citep[e.g.,][]{Yang2020}, the mass density can be estimated as the following
\begin{equation}\label{eq:et}
\rho\approx1.2n_em_p
\end{equation}
The kinetic energy ($E_k$) and thermal energy ($E_t$) can be \textbf{estimated} as
\begin{equation}\label{eq:ek}
E_k=\frac{1}{2}\rho Vv^2\approx 0.6n_em_pL^3v^2
\end{equation}
\begin{equation}\label{eq:et}
E_t=2n_e\frac{3}{2}k_BTV\approx 3n_ek_BTL^3
\end{equation}
where $k_B$ is Boltzmann constant, $\rho$ is the mass density, $V$ is the volume of the blob, and $m_p$ is the proton mass. Here we simply assumed the same length, width and height for the ejected plasma. Considering the projection effect, $E_k$ should be considered as a lower limit. We also listed values of $E_k$ and $E_t$ for all identified blobs in Table~\ref{tab:1}. The average $E_k$ is $10^{22.8\pm0.7}$ erg, and the average $E_t$ is $10^{23.3\pm0.5}$~erg. Both are close to the typical nanoflare energy of $10^{24}$ erg \citep{Parker1988}.

Assuming that the blobs are produced due to the release of magnetic energy, and other energies such as gravitational potential energy and radiation energy are neglegible, the dissipated magnetic energy $E_m$ can be estimated as the sum of $E_k$ and $E_t$ \citep{Priest2014,Chen2015,Chen2020}. Hence, the density of the dissipated magnetic energy can be written as
\begin{equation}
\frac{E_k+E_t}{V}\approx\frac{{(\Delta B)}^2}{8\pi} ,
\end{equation}
where $\Delta B$ represents the dissipated magnetic field. Among the 10 blobs (except blobs 5, 6, and 13), the absolute values of $\Delta B$ are generally 4--6 G. In other words, the magnetic field strength decreases by 4--6 G to trigger the blobs. It should be noted that our estimation based on the assumption of a fully-ionized plasma and a typical coronal composition is not applicable to blob 6 (type \uppercase\expandafter{\romannumeral2}), which could be driven by the magnetic energy release in the lower partially ionized sunspot atmosphere.

\subsection{Possible formation mechanisms\label{subsec:mechanisms}}
We categorized three types of small-scale bright blobs from the LB in Section \ref{subsec:ejections}. \textbf{Due to the lack of spectral information of the blobs, we can only speculate their formation mechanisms from the images.}  The first type is associated with the merging of two bright features on the bright front, and the newly formed features often become brighter. The blobs may be triggered by magnetic reconnection between two approaching plasma blobs. It is also possible that the local magnetic tension strengthens when two brightenings move close to each other, and the enhanced magnetic tension drives these blobs. Both the magnetic reconnection and enhanced magnetic tension could be caused by interaction of adjacent shocks, which leads to the merging of different bright features along the bright fronts.

The second type appears to be related to localized transient brightenings at the base of the LB. These brightenings indicate small-scale magnetic reconnection events on the LB, as suggested by \citet{Hou2017} and \citet{ Tian2018}. Magnetic reconnection between small-scale emerging magnetic bipoles and the surrounding umbral field could occur intermittently and drive upward propagating hot plasma from the LB. The initial acceleration of the blob, which we calculated based on the traveling paths above the bright front, is essentially zero. The negligible acceleration also implies that the origin sites of such blobs are below the bright front, thus the main acceleration takes place before the blobs propagate ahead of the bright front.

The remaining blobs are related to neither merging of bright features on the bright front nor transient brightenings on the LB. For these events, ascending materials from the LB, likely associated with shock waves, are detected to merge with the existing bright front. The shocks may accelerate the local plasma on the bright front, leading to blobs. Merging of plasma at the bright front may also cause instabilities, which might trigger the blobs.

\section{conclusions and discussion \label{sec:conclusion}}
We have performed a detailed study of the kinematics and energetics of small-scale bright blobs above a LB observed in the TR images taken by IRIS. These blobs have a typical lifetime of one minute and an average area of $0.57\pm0.37$ Mm$^2$. All blobs exhibit upward propagation without downward motions. With an average speed of $71.7\pm14.7$ km s$^{-1}$, these blobs could travel to a distance of $\sim$3-7 Mm. Most blobs also appear in the AIA 131, 171, 193, 211, 304, and 335 \AA~images\textbf{ but not in the H$\alpha$ image, suggesting that the blobs are heated above the TR temperatures}. Using the EM-loci method, we found a typical temperature of $10^{5.47\pm0.02}$ K and EM of $10^{27.22\pm0.12}$ cm$^{-5}$ for the ejected plasma. The kinetic energy and thermal energy have been estimated to be $10^{22.8\pm0.7}$ erg and $10^{23.3\pm0.5}$ erg, respectively.

Some blobs occur after the merging of two brightenings on the bright front, or interaction between rising materials and the bright front. These blobs may be triggered by magnetic reconnection, release of locally enhanced magnetic tension, acceleration of shocks, or instabilities at the bright front as shocks travel through. Obvious acceleration has been found for most of these blobs. One blob appears to travel directly from the LB base, and it might be driven by intermittent small-scale reconnection on the LB. This blob reveals a constant velocity, and no obvious acceleration was observed after passing the bright front. \textbf{Though the observations indicate that the bright blobs above the LB are more likely to be mass ejections, it should be noted that we cannot exclude the possibility of the heating fronts.}

Due to the lack of spectral data and magnetic field measurements for these blobs, we cannot provide a conclusive judgement on the origin of these blobs. \textbf{Upcoming 4-m Daniel K. Inouye Solar Telescope (DKIST) would perform unprecedented high-resolution observations of the sunspots and the light bridge and provide simultaneous magnetic field measurements from the photosphere to the chromosphere. The Spectral Imaging of the Coronal Environment (SPICE) onboard Solar Orbiter could provide the spectra of the lines with a wide range of formation temperatures. These new facilities will likely provide new insight into the generation mechanisms of these bright blobs.} Advanced three-dimensional numerical simulations of active regions, including dynamics above light bridges, will also help us better understand their formation mechanisms.

\acknowledgments
This work is supported by NSFC grants 11825301, 11803002, 11790304(11790300), 11427901, 11873062, 11903050, and 11773039 and the Strategic Priority Research Program of CAS (grant XDA17040507). IRIS is a NASA Small Explorer mission developed and operated by LMSAL with mission operations executed at NASA Ames Research center and major contributions to downlink communications funded by ESA and the Norwegian Space Center. SDO is a space mission in the Living With a Star Program of NASA.

\vspace{5mm}

\bibliography{sample}{}

\begin{thebibliography}{}
\expandafter\ifx\csname natexlab\endcsname\relax\def\natexlab#1{#1}\fi
\providecommand{\url}[1]{\href{#1}{#1}}
\providecommand{\dodoi}[1]{doi:~\href{http://doi.org/#1}{\nolinkurl{#1}}}
\providecommand{\doeprint}[1]{\href{http://ascl.net/#1}{\nolinkurl{http://ascl.net/#1}}}
\providecommand{\doarXiv}[1]{\href{https://arxiv.org/abs/#1}{\nolinkurl{https://arxiv.org/abs/#1}}}

\bibitem[{{Asai} {et~al.}(2001){Asai}, {Ishii}, \& {Kurokawa}}]{Asai2001}
{Asai}, A., {Ishii}, T.~T., \& {Kurokawa}, H. 2001, \apjl, 555, L65,
  \dodoi{10.1086/321738}

\bibitem[{{Bai} {et~al.}(2019){Bai}, {Socas-Navarro}, {N{\'o}brega-Siverio},
  {Su}, {Deng}, {Li}, {Cao}, \& {Ji}}]{Bai2019}
{Bai}, X., {Socas-Navarro}, H., {N{\'o}brega-Siverio}, D., {et~al.} 2019, \apj,
  870, 90, \dodoi{10.3847/1538-4357/aaf1d1}

\bibitem[{{Berger} \& {Berdyugina}(2003)}]{Berger2003}
{Berger}, T.~E., \& {Berdyugina}, S.~V. 2003, \apjl, 589, L117,
  \dodoi{10.1086/376494}

\bibitem[{{Bharti}(2015)}]{Bharti2015}
{Bharti}, L. 2015, \mnras, 452, L16, \dodoi{10.1093/mnrasl/slv071}

\bibitem[{{Bharti} {et~al.}(2007){Bharti}, {Rimmele}, {Jain}, {Jaaffrey}, \&
  {Smartt}}]{Bharti2007}
{Bharti}, L., {Rimmele}, T., {Jain}, R., {Jaaffrey}, S.~N.~A., \& {Smartt},
  R.~N. 2007, \mnras, 376, 1291, \dodoi{10.1111/j.1365-2966.2007.11525.x}

\bibitem[{{Cao} {et~al.}(2010){Cao}, {Gorceix}, {Coulter}, {Ahn}, {Rimmele}, \&
  {Goode}}]{Cao2010}
{Cao}, W., {Gorceix}, N., {Coulter}, R., {et~al.} 2010, Astronomische
  Nachrichten, 331, 636, \dodoi{10.1002/asna.201011390}

\bibitem[{{Chen} {et~al.}(2020){Chen}, {Zhang}, {De Pontieu}, {Ma}, {Kliem}, \&
  {Priest}}]{Chen2020}
{Chen}, H., {Zhang}, J., {De Pontieu}, B., {et~al.} 2020, \apj, 899, 19,
  \dodoi{10.3847/1538-4357/ab9cad}

\bibitem[{{Chen} {et~al.}(2015){Chen}, {Su}, {Yin}, {Priya}, {Zhang}, {Liu},
  {Xu}, \& {Yu}}]{Chen2015}
{Chen}, J., {Su}, J., {Yin}, Z., {et~al.} 2015, \apj, 815, 71,
  \dodoi{10.1088/0004-637X/815/1/71}

\bibitem[{{Chen} {et~al.}(2019){Chen}, {Tian}, {Huang}, {Peter}, \&
  {Samanta}}]{Chen2019}
{Chen}, Y., {Tian}, H., {Huang}, Z., {Peter}, H., \& {Samanta}, T. 2019, \apj,
  873, 79, \dodoi{10.3847/1538-4357/ab0417}

\bibitem[{{De Pontieu} {et~al.}(2017){De Pontieu}, {Mart{\'\i}nez-Sykora}, \&
  {Chintzoglou}}]{Pontieu2017}
{De Pontieu}, B., {Mart{\'\i}nez-Sykora}, J., \& {Chintzoglou}, G. 2017, \apjl,
  849, L7, \dodoi{10.3847/2041-8213/aa9272}

\bibitem[{{De Pontieu} {et~al.}(2014){De Pontieu}, {Title}, {Lemen}, {Kushner},
  {Akin}, {Allard}, {Berger}, {Boerner}, {Cheung}, {Chou}, {Drake}, {Duncan},
  {Freeland}, {Heyman}, {Hoffman}, {Hurlburt}, {Lindgren}, {Mathur}, {Rehse},
  {Sabolish}, {Seguin}, {Schrijver}, {Tarbell}, {W{\"u}lser}, {Wolfson},
  {Yanari}, {Mudge}, {Nguyen-Phuc}, {Timmons}, {van Bezooijen}, {Weingrod},
  {Brookner}, {Butcher}, {Dougherty}, {Eder}, {Knagenhjelm}, {Larsen},
  {Mansir}, {Phan}, {Boyle}, {Cheimets}, {DeLuca}, {Golub}, {Gates}, {Hertz},
  {McKillop}, {Park}, {Perry}, {Podgorski}, {Reeves}, {Saar}, {Testa}, {Tian},
  {Weber}, {Dunn}, {Eccles}, {Jaeggli}, {Kankelborg}, {Mashburn}, {Pust},
  {Springer}, {Carvalho}, {Kleint}, {Marmie}, {Mazmanian}, {Pereira}, {Sawyer},
  {Strong}, {Worden}, {Carlsson}, {Hansteen}, {Leenaarts}, {Wiesmann},
  {Aloise}, {Chu}, {Bush}, {Scherrer}, {Brekke}, {Martinez-Sykora}, {Lites},
  {McIntosh}, {Uitenbroek}, {Okamoto}, {Gummin}, {Auker}, {Jerram}, {Pool}, \&
  {Waltham}}]{De2014}
{De Pontieu}, B., {Title}, A.~M., {Lemen}, J.~R., {et~al.} 2014, \solphys, 289,
  2733, \dodoi{10.1007/s11207-014-0485-y}

\bibitem[{{Del Zanna} {et~al.}(2002){Del Zanna}, {Landini}, \&
  {Mason}}]{Del2002}
{Del Zanna}, G., {Landini}, M., \& {Mason}, H.~E. 2002, \aap, 385, 968,
  \dodoi{10.1051/0004-6361:20020164}

\bibitem[{{Del Zanna} {et~al.}(2011){Del Zanna}, {O'Dwyer}, \&
  {Mason}}]{Del2011}
{Del Zanna}, G., {O'Dwyer}, B., \& {Mason}, H.~E. 2011, \aap, 535, A46,
  \dodoi{10.1051/0004-6361/201117470}

\bibitem[{{Esteban Pozuelo} {et~al.}(2019){Esteban Pozuelo}, {de la Cruz
  Rodr{\'\i}guez}, {Drews}, {Rouppe van der Voort}, {Scharmer}, \&
  {Carlsson}}]{Esteban2019}
{Esteban Pozuelo}, S., {de la Cruz Rodr{\'\i}guez}, J., {Drews}, A., {et~al.}
  2019, \apj, 870, 88, \dodoi{10.3847/1538-4357/aaf28a}

\bibitem[{{Feng} {et~al.}(2020){Feng}, {Miao}, {Yuan}, {Qu}, \&
  {Nakariakov}}]{Feng2020}
{Feng}, S., {Miao}, Y., {Yuan}, D., {Qu}, Z., \& {Nakariakov}, V.~M. 2020,
  \apjl, 893, L2, \dodoi{10.3847/2041-8213/ab7dc4}

\bibitem[{{Hou} {et~al.}(2016{\natexlab{a}}){Hou}, {Zhang}, {Li}, {Yang}, {Li},
  \& {Li}}]{Hou2016a}
{Hou}, Y., {Zhang}, J., {Li}, T., {et~al.} 2016{\natexlab{a}}, \apjl, 829, L29,
  \dodoi{10.3847/2041-8205/829/2/L29}

\bibitem[{{Hou} {et~al.}(2017){Hou}, {Zhang}, {Li}, {Yang}, \& {Li}}]{Hou2017}
{Hou}, Y., {Zhang}, J., {Li}, T., {Yang}, S., \& {Li}, X. 2017, \apjl, 848, L9,
  \dodoi{10.3847/2041-8213/aa8edd}

\bibitem[{{Hou} {et~al.}(2016{\natexlab{b}}){Hou}, {Li}, {Yang}, \&
  {Zhang}}]{Hou2016b}
{Hou}, Y.~J., {Li}, T., {Yang}, S.~H., \& {Zhang}, J. 2016{\natexlab{b}}, \aap,
  589, L7, \dodoi{10.1051/0004-6361/201628216}

\bibitem[{{Leka}(1997)}]{Leka1997}
{Leka}, K.~D. 1997, \apj, 484, 900, \dodoi{10.1086/304363}

\bibitem[{{Lemen} {et~al.}(2012){Lemen}, {Title}, {Akin}, {Boerner}, {Chou},
  {Drake}, {Duncan}, {Edwards}, {Friedlaender}, {Heyman}, {Hurlburt}, {Katz},
  {Kushner}, {Levay}, {Lindgren}, {Mathur}, {McFeaters}, {Mitchell}, {Rehse},
  {Schrijver}, {Springer}, {Stern}, {Tarbell}, {Wuelser}, {Wolfson}, {Yanari},
  {Bookbinder}, {Cheimets}, {Caldwell}, {Deluca}, {Gates}, {Golub}, {Park},
  {Podgorski}, {Bush}, {Scherrer}, {Gummin}, {Smith}, {Auker}, {Jerram},
  {Pool}, {Soufli}, {Windt}, {Beardsley}, {Clapp}, {Lang}, \&
  {Waltham}}]{Lemen2012}
{Lemen}, J.~R., {Title}, A.~M., {Akin}, D.~J., {et~al.} 2012, \solphys, 275,
  17, \dodoi{10.1007/s11207-011-9776-8}

\bibitem[{{Lim} {et~al.}(2020){Lim}, {Yang}, {Yurchyshyn}, {Chae}, {Song}, \&
  {Madjarska}}]{Lim2020}
{Lim}, E.-K., {Yang}, H., {Yurchyshyn}, V., {et~al.} 2020, arXiv e-prints,
  arXiv:2010.10713.
\newblock \doarXiv{2010.10713}

\bibitem[{{Lim} {et~al.}(2011){Lim}, {Yurchyshyn}, {Abramenko}, {Ahn}, {Cao},
  \& {Goode}}]{Lim2011}
{Lim}, E.-K., {Yurchyshyn}, V., {Abramenko}, V., {et~al.} 2011, \apj, 740, 82,
  \dodoi{10.1088/0004-637X/740/2/82}

\bibitem[{{Liu} {et~al.}(2014){Liu}, {Xu}, {Gu}, {Wang}, {You}, {Shen}, {Lu},
  {Jin}, {Chen}, {Lou}, {Li}, {Liu}, {Xu}, {Rao}, {Hu}, {Li}, {Fu}, {Wang},
  {Bao}, {Wu}, \& {Zhang}}]{Liu2014}
{Liu}, Z., {Xu}, J., {Gu}, B.-Z., {et~al.} 2014, Research in Astronomy and
  Astrophysics, 14, 705, \dodoi{10.1088/1674-4527/14/6/009}

\bibitem[{{Louis} {et~al.}(2014){Louis}, {Beck}, \& {Ichimoto}}]{Louis2014}
{Louis}, R.~E., {Beck}, C., \& {Ichimoto}, K. 2014, \aap, 567, A96,
  \dodoi{10.1051/0004-6361/201423756}

\bibitem[{{Louis} {et~al.}(2015){Louis}, {Bellot Rubio}, {de la Cruz
  Rodr{\'\i}guez}, {Socas-Navarro}, \& {Ortiz}}]{Louis2015}
{Louis}, R.~E., {Bellot Rubio}, L.~R., {de la Cruz Rodr{\'\i}guez}, J.,
  {Socas-Navarro}, H., \& {Ortiz}, A. 2015, \aap, 584, A1,
  \dodoi{10.1051/0004-6361/201526854}

\bibitem[{{Morton}(2012)}]{Morton2012}
{Morton}, R.~J. 2012, \aap, 543, A6, \dodoi{10.1051/0004-6361/201219137}

\bibitem[{{Parker}(1988)}]{Parker1988}
{Parker}, E.~N. 1988, \apj, 330, 474, \dodoi{10.1086/166485}

\bibitem[{{Priest}(2014)}]{Priest2014}
{Priest}, E. 2014, {Magnetohydrodynamics of the Sun},
  \dodoi{10.1017/CBO9781139020732}

\bibitem[{{Robustini} {et~al.}(2016){Robustini}, {Leenaarts}, {de la Cruz
  Rodriguez}, \& {Rouppe van der Voort}}]{Robustini2016}
{Robustini}, C., {Leenaarts}, J., {de la Cruz Rodriguez}, J., \& {Rouppe van
  der Voort}, L. 2016, \aap, 590, A57, \dodoi{10.1051/0004-6361/201528022}

\bibitem[{{Rouppe van der Voort} {et~al.}(2010){Rouppe van der Voort}, {Bellot
  Rubio}, \& {Ortiz}}]{Rouppe2010}
{Rouppe van der Voort}, L., {Bellot Rubio}, L.~R., \& {Ortiz}, A. 2010, \apjl,
  718, L78, \dodoi{10.1088/2041-8205/718/2/L78}

\bibitem[{{Rouppe van der Voort} \& {Drews}(2019)}]{Rouppe2019}
{Rouppe van der Voort}, L. H.~M., \& {Drews}, A. 2019, \aap, 626, A62,
  \dodoi{10.1051/0004-6361/201935343}

\bibitem[{{Roy}(1973)}]{Roy1973}
{Roy}, J.~R. 1973, \solphys, 28, 95, \dodoi{10.1007/BF00152915}

\bibitem[{{Sch{\"u}ssler} \& {V{\"o}gler}(2006)}]{Sch2006}
{Sch{\"u}ssler}, M., \& {V{\"o}gler}, A. 2006, \apjl, 641, L73,
  \dodoi{10.1086/503772}

\bibitem[{{Shimizu} {et~al.}(2009){Shimizu}, {Katsukawa}, {Kubo}, {Lites},
  {Ichimoto}, {Suematsu}, {Tsuneta}, {Nagata}, {Shine}, \&
  {Tarbell}}]{Shimizu2009}
{Shimizu}, T., {Katsukawa}, Y., {Kubo}, M., {et~al.} 2009, \apjl, 696, L66,
  \dodoi{10.1088/0004-637X/696/1/L66}

\bibitem[{{Song} {et~al.}(2017){Song}, {Chae}, {Yurchyshyn}, {Lim}, {Cho},
  {Yang}, {Cho}, \& {Kwak}}]{Song2017}
{Song}, D., {Chae}, J., {Yurchyshyn}, V., {et~al.} 2017, \apj, 835, 240,
  \dodoi{10.3847/1538-4357/835/2/240}

\bibitem[{{Testa} {et~al.}(2012){Testa}, {De Pontieu}, {Mart{\'\i}nez-Sykora},
  {Hansteen}, \& {Carlsson}}]{Testa2012}
{Testa}, P., {De Pontieu}, B., {Mart{\'\i}nez-Sykora}, J., {Hansteen}, V., \&
  {Carlsson}, M. 2012, \apj, 758, 54, \dodoi{10.1088/0004-637X/758/1/54}

\bibitem[{{Tian} {et~al.}(2014){Tian}, {Kleint}, {Peter}, {Weber}, {Testa},
  {DeLuca}, {Golub}, \& {Schanche}}]{Tian2014}
{Tian}, H., {Kleint}, L., {Peter}, H., {et~al.} 2014, \apjl, 790, L29,
  \dodoi{10.1088/2041-8205/790/2/L29}

\bibitem[{{Tian} {et~al.}(2018{\natexlab{a}}){Tian}, {Zhu}, {Peter}, {Zhao},
  {Samanta}, \& {Chen}}]{Tian2018a}
{Tian}, H., {Zhu}, X., {Peter}, H., {et~al.} 2018{\natexlab{a}}, \apj, 854,
  174, \dodoi{10.3847/1538-4357/aaaae6}

\bibitem[{{Tian} {et~al.}(2018{\natexlab{b}}){Tian}, {Yurchyshyn}, {Peter},
  {Solanki}, {Young}, {Ni}, {Cao}, {Ji}, {Zhu}, {Zhang}, {Samanta}, {Song},
  {He}, {Wang}, \& {Chen}}]{Tian2018}
{Tian}, H., {Yurchyshyn}, V., {Peter}, H., {et~al.} 2018{\natexlab{b}}, \apj,
  854, 92, \dodoi{10.3847/1538-4357/aaa89d}

\bibitem[{{Toriumi} {et~al.}(2015{\natexlab{a}}){Toriumi}, {Cheung}, \&
  {Katsukawa}}]{Toriumi2015b}
{Toriumi}, S., {Cheung}, M. C.~M., \& {Katsukawa}, Y. 2015{\natexlab{a}}, \apj,
  811, 138, \dodoi{10.1088/0004-637X/811/2/138}

\bibitem[{{Toriumi} {et~al.}(2015{\natexlab{b}}){Toriumi}, {Katsukawa}, \&
  {Cheung}}]{Toriumi2015a}
{Toriumi}, S., {Katsukawa}, Y., \& {Cheung}, M. C.~M. 2015{\natexlab{b}}, \apj,
  811, 137, \dodoi{10.1088/0004-637X/811/2/137}

\bibitem[{{Winebarger} {et~al.}(2013){Winebarger}, {Walsh}, {Moore}, {De
  Pontieu}, {Hansteen}, {Cirtain}, {Golub}, {Kobayashi}, {Korreck}, {DeForest},
  {Weber}, {Title}, \& {Kuzin}}]{Amy2013}
{Winebarger}, A.~R., {Walsh}, R.~W., {Moore}, R., {et~al.} 2013, \apj, 771, 21,
  \dodoi{10.1088/0004-637X/771/1/21}

\bibitem[{{Yan} {et~al.}(2020){Yan}, {Liu}, {Zhang}, \& {Xu}}]{Yan2020}
{Yan}, X., {Liu}, Z., {Zhang}, J., \& {Xu}, Z. 2020, Science in China E:
  Technological Sciences, 63, 1656, \dodoi{10.1007/s11431-019-1463-6}

\bibitem[{{Yang} {et~al.}(2019{\natexlab{a}}){Yang}, {Lim}, {Iijima},
  {Yurchyshyn}, {Cho}, {Lee}, {Schmieder}, {Kim}, {Kim}, \&
  {Bong}}]{YangHeesu2019}
{Yang}, H., {Lim}, E.-K., {Iijima}, H., {et~al.} 2019{\natexlab{a}}, \apj, 882,
  175, \dodoi{10.3847/1538-4357/ab36b7}

\bibitem[{{Yang} {et~al.}(2016){Yang}, {Zhang}, \& {Erd{\'e}lyi}}]{Yang2016}
{Yang}, S., {Zhang}, J., \& {Erd{\'e}lyi}, R. 2016, \apjl, 833, L18,
  \dodoi{10.3847/2041-8213/833/2/L18}

\bibitem[{{Yang} {et~al.}(2015){Yang}, {Zhang}, {Jiang}, \& {Xiang}}]{Yang2015}
{Yang}, S., {Zhang}, J., {Jiang}, F., \& {Xiang}, Y. 2015, \apjl, 804, L27,
  \dodoi{10.1088/2041-8205/804/2/L27}

\bibitem[{{Yang} {et~al.}(2019{\natexlab{b}}){Yang}, {Yurchyshyn}, {Ahn},
  {Penn}, \& {Cao}}]{YangXu2019}
{Yang}, X., {Yurchyshyn}, V., {Ahn}, K., {Penn}, M., \& {Cao}, W.
  2019{\natexlab{b}}, \apj, 886, 64, \dodoi{10.3847/1538-4357/ab4a7d}

\bibitem[{{Yang} {et~al.}(2020){Yang}, {Bethge}, {Tian}, {Tomczyk}, {Morton},
  {Del Zanna}, {McIntosh}, {Karak}, {Gibson}, {Samanta}, {He}, {Chen}, \&
  {Wang}}]{Yang2020}
{Yang}, Z., {Bethge}, C., {Tian}, H., {et~al.} 2020, Science, 369, 694,
  \dodoi{10.1126/science.abb4462}

\bibitem[{{Yuan} \& {Walsh}(2016)}]{Yuan2016}
{Yuan}, D., \& {Walsh}, R.~W. 2016, \aap, 594, A101,
  \dodoi{10.1051/0004-6361/201629258}

\bibitem[{{Zhang} {et~al.}(2017){Zhang}, {Tian}, {He}, \& {Wang}}]{Zhang2017}
{Zhang}, J., {Tian}, H., {He}, J., \& {Wang}, L. 2017, \apj, 838, 2,
  \dodoi{10.3847/1538-4357/aa63e8}

\bibitem[{{Zhang} {et~al.}(2018){Zhang}, {Tian}, {Solanki}, {Wang}, {Peter},
  {Ahn}, {Xu}, {Zhu}, {Cao}, {He}, \& {Wang}}]{Zhang2018}
{Zhang}, J., {Tian}, H., {Solanki}, S.~K., {et~al.} 2018, \apj, 865, 29,
  \dodoi{10.3847/1538-4357/aada0a}

\end{thebibliography}
\bibliographystyle{aasjournal}

\end{document}